\documentclass[sigconf,natbib=false]{acmart}
\AtBeginDocument{%
  }

\usepackage{natbib}
\setcitestyle{square,numbers,sort&compress}
\bibpunct{[}{]}{,}{n}{}{}
\usepackage{balance}
\usepackage{subcaption}
\usepackage[table,xcdraw]{xcolor}
\usepackage{graphicx}
\usepackage{booktabs}
\usepackage{relsize}
\usepackage{colortbl}
\usepackage{enumerate}
\usepackage{tikz}
\usepackage{pgf}
\usepackage{pbox}
\usepackage{rotating}
\usepackage{multirow}
\usepackage{lipsum}
\usepackage{tablefootnote}
\usepackage{cleveref}
\crefformat{footnote}{#2\footnotemark[#1]#3}
\usepackage{soul}

\usepackage{url}
\usepackage{algpseudocode}
\usepackage{listings}
\usepackage{float}
\usepackage{textcomp}

\newcommand{\startlist}{\begin{list}{\labelitemi}{\leftmargin=1em}\setlength{\itemsep}{-1mm}}
\newcommand{\stoplist}{\end{list}}

\newcommand{\rqone}{How well do RovoDev-generated comments align with human-written comments?}

\newcommand{\rqfour}{How frequently do software engineers resolved the comments generated by RovoDev Code Reviewers when compared to human-written comments?}

\newcommand{\rqfive}{How does the adoption of RovoDev-generated comments impact the code review workflow?}

\newcommand{\rqsix}{How do software engineers perceive the quality of the code review comments generated by RovoDev Code Reviewer?}

\newcommand*\circled[1]{\tikz[baseline=(char.base)]{
            \node[shape=circle,draw,inner sep=0.5pt] (char) {\small{#1}};}}

\newcommand{\smallsection}[1]{\noindent {\bf #1}.\hspace{1mm}}

\copyrightyear{2026}
\acmYear{2026}
\setcopyright{cc}
\setcctype{by-nc-nd}
\acmConference[ICSE-SEIP '26]{2026 IEEE/ACM 48th International Conference on Software Engineering}{April 12--18, 2026}{Rio de Janeiro, Brazil}
\acmBooktitle{2026 IEEE/ACM 48th International Conference on Software Engineering (ICSE-SEIP '26), April 12--18, 2026, Rio de Janeiro, Brazil}
\acmPrice{}
\acmDOI{10.1145/3786583.3786851}
\acmISBN{979-8-4007-2426-8/2026/04}

\begin{document}

\title{RovoDev Code Reviewer: A Large-Scale Online Evaluation of LLM-based Code Review Automation at Atlassian}



\author{Kla Tantithamthavorn}
\affiliation{%
  \institution{Monash University \& Atlassian}
  \country{Australia.}
}

\author{Yaotian Zou}
\affiliation{%
  \institution{Atlassian}
  \country{USA.}
}

\author{Andy Wong}
\affiliation{%
  \institution{Atlassian}
  \country{USA.}
}

\author{Michael Gupta}
\affiliation{%
  \institution{Atlassian}
  \country{USA.}
}

\author{Zhe Wang}
\affiliation{%
  \institution{Atlassian}
  \country{Australia.}
}

\author{Mike Buller}
\affiliation{%
  \institution{Atlassian}
  \country{Australia.}
}

\author{Ryan Jiang}
\affiliation{%
  \institution{Atlassian}
  \country{Australia.}
}

\author{Matthew Watson}
\affiliation{%
  \institution{Atlassian}
  \country{Australia.}
}

\author{Minwoo Jeong}
\affiliation{%
  \institution{Atlassian}
  \country{USA.}
}

\author{Kun Chen}
\affiliation{%
  \institution{Atlassian}
  \country{USA.}
}

\author{Ming Wu}
\affiliation{%
  \institution{Atlassian}
  \country{USA.}
}


\renewcommand{\shortauthors}{Tantithamthavorn et al.}

\begin{abstract}
Large Language Models (LLMs)-powered code review automation has the potential to transform code review workflows.
Despite the advances of LLM-powered code review comment generation approaches, several practical challenges remain for designing enterprise-grade code review automation tools. 
In particular, this paper aims at answering the practical question: \emph{how can we design a review-guided, context-aware, quality-checked code review comment generation without fine-tuning?}
In this paper, we present RovoDev Code Reviewer, an enterprise-grade LLM-based code review automation tool designed and deployed at scale within Atlassian's development ecosystem with seamless integration into Atlassian's Bitbucket. 
Through the offline, online, user feedback evaluations over a one-year period, we conclude that RovoDev Code Reviewer is effective in generating code review comments that could lead to code resolution for 38.70\% (i.e., comments that triggered code changes in the subsequent commits); and offers the promise of accelerating feedback cycles (i.e., decreasing the PR cycle time by 30.8\%), alleviating reviewer workload (i.e., reducing the number of human-written comments by 35.6\%), and improving overall software quality (i.e., finding errors with actionable suggestions).
\end{abstract}




\maketitle

\section{Introduction}

Code review is a cornerstone of modern software engineering~\cite{bacchelli2013expectations,rigby2011broadcast, rigby2013decade, macleod2017code}, serving as a critical quality assurance practice that helps teams identify defects, share knowledge, and maintain high coding standards. 
However, as software projects grow in size and complexity, manual code review becomes increasingly time-consuming and resource-intensive~\cite{thongtanunam2015should}, often leading to bottlenecks in the development process.
Automating aspects of code review using advances in Large Language Models (LLMs)~\cite{tufano2025automating,tufano2024code,tufano2019learning,thongtanunam2014improving,thongtanunam2022autotransform,hong2022commentfinder,hong2022should} could speed up code review processes, particularly, \emph{code review comment generation}, defined as a generative task to generate code review comments written in natural languages for a given code change~\cite{hong2022commentfinder,frommgen2024resolving,hong2025retrieval,olewicki2024impact,tufano2022using}.

Despite these advances, several practical challenges remain for designing enterprise-grade code review automation tools. 
\emph{First}, data privacy and security are paramount, especially when processing sensitive customer code and metadata, making it infeasible to fine-tune LLMs on proprietary or user-generated content in many industrial contexts. 
\emph{Second}, review guidelines play a critical role to guide inexperienced reviewers to conduct a code review, yet it remains largely ignored in the recent LLM-powered code review comment generation approaches~\cite{olewicki2024impact}.
\emph{Third}, most retrieval-augmented generation (RAG) approaches~\cite{olewicki2024impact, hong2025retrieval} rely on rich historical data, which may not be available for newly created or context-limited projects. 
\emph{Finally}, LLMs are prone to generating noisy or hallucinated comments that may be vague, non-actionable, or factually incorrect~\cite{liu2025too,she2023pitfalls}, potentially diminishing the utility of automated code review tools.

\emph{In this paper}, we present RovoDev Code Reviewer, a  Review-Guided, Quality-Checked Code Review Automation.
Our RovoDev Code Reviewer consists of three key components: (1) a zero-shot context-aware review-guided comment generation; (2) a comment quality check on factual correctness to remove noisy comments (e.g., irrelevant, inaccurate, inconsistent, or nonsensical); and (3) a comment quality check on actionability to ensure that the RovoDev-generated comments are most likely to lead to code resolution (i.e., the code lines being commented are resolved in the next commit). 

Similar to traditional code review benchmark datasets, the offline evaluation aspects are limited to the human alignment (i.e., whether an LLM comment is semantically similar to the given human comment), are static (no interaction with software engineers), and do not capture the full complexity or evolving nature of real-world software engineering tasks.
In reality, code review comments can be written in many different forms, regardless of their intended purpose. 
To better capture the utility of RovoDev Code Reviewer,  we implemented and deployed across more than 1,900 Atlassian's source code repositories, resulting in the generation of over 54,000 review comments.
Then, we rigorously assess its practical value and its impact on Atlassian's internal code review workflow, which goes far beyond typical academic evaluations that rely solely on static datasets.
Below, we answer the following three RQs:


\begin{enumerate}[{\bf RQ1)}]
\item {\bf \rqfour}\\
RovoDev Code Reviewer achieves a code resolution rate of 38.70\%, which is 12.9\% relatively less than a code resolution rate of human-written comments (44.45\%).

\item {\bf \rqfive}\\
RovoDev has a significant positive impact on the code review workflow at Altlassian, accelerating the median PR cycle time by 30.8\%, and reducing the number of human-written comments by 35.6\%.

\item {\bf \rqsix}\\
Software engineers found that RovoDev can provide comments with accurate error detection and actionable suggestions.
However, RovoDev may generate incorrect or non-actionable comments when the context information is unknown.

\end{enumerate}

Based on our year-long large-scale online production and evaluation with over 1,900 repositories, we conclude that RovoDev Code Reviewer is effective in generating code review comments that could lead to code resolution (i.e., comments that triggered code changes in the subsequent commits), offers positive impact on code review workflows (i.e., decreasing the PR cycle time, and reducing the number of the human-written comments), with positive qualitative feedback from several practitioners (i.e., finding errors with actionable suggestions).

\begin{figure*}[t]
\centering
\includegraphics[width=.85\linewidth]{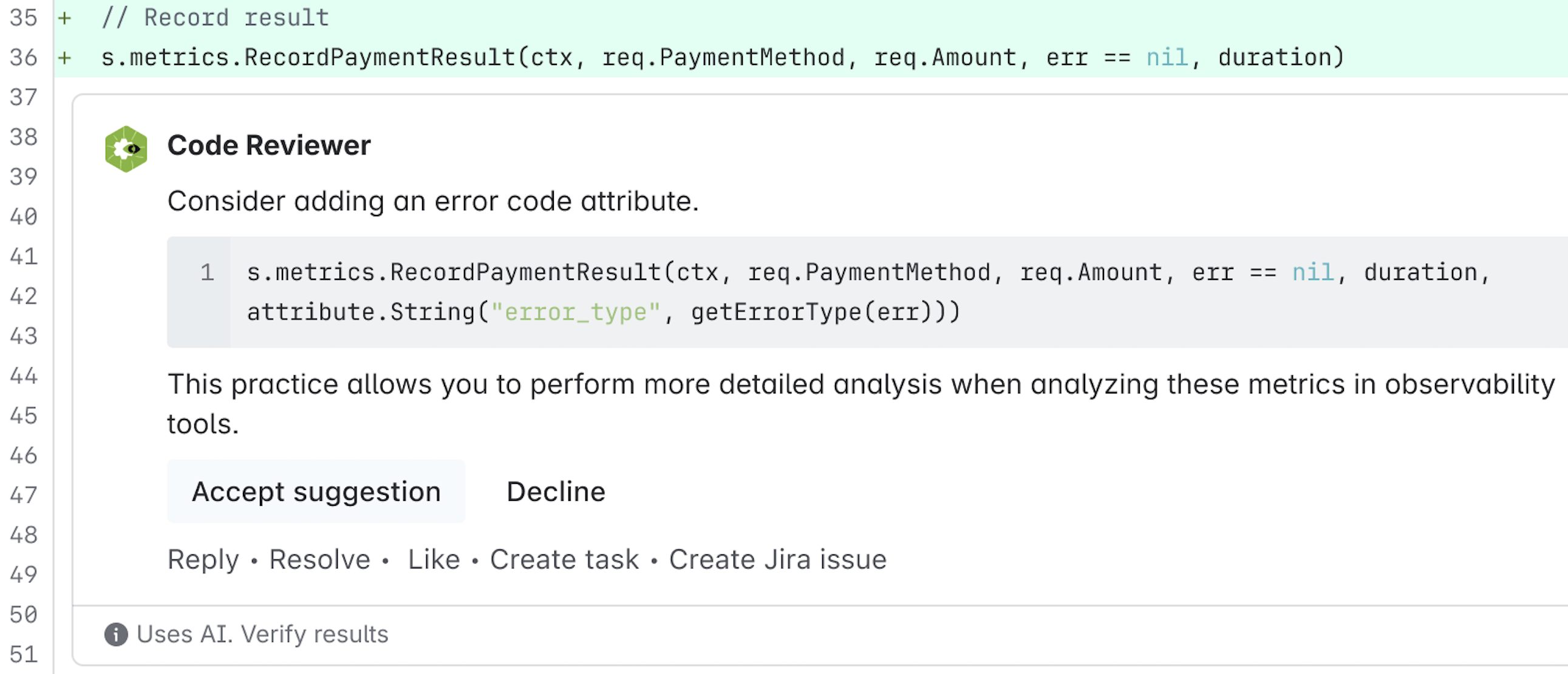}
\caption{An example of a code review comment generated by RovoDev Code Reviewer.}
\label{fig:rovodevcodereviewer}
\end{figure*}


\textbf{Novelty \& Contributions.} This paper is the first to present:

\begin{enumerate}
    \item RovoDev Code Reviewer: A Context-Aware Review-Guided Quality-Checked LLM-based Code Review Automation (Section 3).
    \item A year-long online deployment and evaluation of RovoDev Code Reviewer across more than 1,900 source code repositories, resulting in the generation of over 54,000 code review comments (RQ1 to RQ3).
\end{enumerate}

\textbf{Paper Organization.}
The paper is organized as follows.
Section~\ref{sec:background} discusses Modern Code Review at Atlassian, related works, and limitations.
Section~\ref{sec:rovodev} presents Atlassian's RovoDev Code Reviewer. 
Section~\ref{sec:experiment} presents the experimental design and the results of our offline and online evaluation.
Section~\ref{sec:threats} discloses the threats to validity.
Finally, Section~\ref{sec:conclusion} draws the conclusion. 


\section{Background and Related Work}\label{sec:background}

In this section, we briefly describe modern code review at Atlassian, the need of code review automation in an industrial context, and the limitations of prior work.

\subsection{Modern Code Review at Atlassian}

Modern code review (MCR) has become a cornerstone of contemporary software engineering practice~\cite{rigby2011broadcast,rigby2013decade}, replacing traditional, formal inspections with lightweight, tool-supported, and asynchronous processes. 
Unlike earlier approaches that relied on in-person meetings and checklists, MCR is typically conducted using platforms such as  Bitbucket~\cite{bitbucket}, Gerrit~\cite{gerrit}, GitHub~\cite{github}, GitLab~\cite{gitlab}, which facilitate distributed and collaborative review of code changes~\cite{bacchelli2013expectations}.

\textbf{At Atlassian} (a global technology company known for its suite of software development tools, such as, Bitbucket, Jira, and Confluence)~\cite{atlassian}, code review is a mandatory and deeply embedded practice across all software engineering teams.
Every code change, regardless of size, is subject to peer review before it can be merged into the main codebase, ensuring its code quality, maintainability, and security.
This process is facilitated primarily through pull requests in Bitbucket
, where software engineers submit their changes and request feedback from reviewers.
The process is collaborative, with reviewers and authors engaging in discussions to clarify intent, resolve issues, and reach consensus before changes are approved.

\textbf{Challenges.} Given the scale and rapid pace of development at Atlassian, code review can become a bottleneck, especially during periods of high activity or when teams are distributed across multiple time zones. 
Delays in review cycles can slow down feature delivery and impact engineers' productivity and the code review workflow. 
However, the core review process remains largely manual, relying on the expertise and availability of human reviewers.

\subsection{Large Language Models-based Code Review Automation and Limitations}

Code review automation~\cite{tufano2025automating} has been an active area of research, driven by the need to improve software quality, engineers' productivity, and the code review speed. 
With the rise of machine learning and natural language processing, researchers have explored data-driven approaches to code review automation~\cite{li2022automating}.
For example, reviewer recommendation~\cite{balachandran2013reducing,thongtanunam2015should,thongtanunam2014improving,hannebauer2016automatically}, code review prioritization~\cite{fu2022linevul,pornprasit2021jitline,wattanakriengkrai2020predicting}, line-level localization~\cite{hong2022should}, code review comment generation~\cite{hong2025retrieval, frommgen2024resolving, olewicki2024impact, liu2025too} and recommendation~\cite{hong2022commentfinder}, code review usefulness~\cite{bosu2015characteristics,pangsakulyanont2014assessing}, and code refinement~\cite{thongtanunam2022autotransform, tufan2021towards, tufano2019learning, fu2022VulRepair, lu2023llama}.

\textbf{LLM-driven code review comment generation} is an emerging research area that leverages Large Language Models (LLMs) to automatically generate review comments for a given code change.
Recent studies have explored the use of off-the-shelf LLMs, such as, Neural Machine Translation~\cite{tufan2021towards}, T5~\cite{tufano2022using}, CodeT5~\cite{tufano2024code}, Llama~\cite{llamareviewer}, GPT-3.5~\cite{pornprasit2024fine}, and GPT-4o~\cite{olewicki2024impact}.
Many studies have demonstrated that fine-tuning LLMs on task-specific data yields the best effectiveness for code review comment generation. 
This process typically relies on datasets consisting of code and corresponding human-written comments, often sourced from publicly available repositories (e.g., TufanoT5~\cite{tufan2021towards}). 
However, several practical yet unsolved challenges arise when designing an enterprise-grade code review comment generation tool.

\textbf{First, an enterprise-grade code review comment generation tool should protect customers' code and related metadata}.
For Atlassian's enterprise customers, data privacy and security are top concerns—nearly half cite these as critical blockers to adopting AI-powered software development tools~\cite{torgerson2024datasecurity}.
Mishandling customer data can lead to severe consequences~\cite{atlassian2025privacy}, including breaches of confidentiality, loss of customer trust, regulatory non-compliance (such as GDPR or CCPA), and significant financial or reputation damage.
As a result, fine-tuning large language models, which is common in the existing work~\cite{hong2025retrieval, tufano2021towards}, for code review comment generation may pose a risk of data security and privacy for enterprise-grade automated code review tools. 

\textbf{Second, an enterprise-grade code review comment generation tool should be equipped with high-standard code review guidelines.} 
Code review guidelines or checklists often serve as a fundamental component of code review processes.
Prior study found that checklist-based code review is an effective mechanism for helping inexperienced reviewers to conduct code review processes, since the checklist can guide them to focus on the important issues or defects when performing code reviews~\cite{chong2021assessing}.
However, such code review checklists have been ignored in the recent automated code review approaches~\cite{olewicki2024impact}.

\textbf{Third, an enterprise-grade code review comment generation tool should be context-aware, while being capable of supporting newly created software projects with limited contextual information.}
Recent work proposed to use Retrieval-Augmented Generation (RAG)~\cite{hong2025retrieval, olewicki2024impact} to retrieve within-repository contextual information to further improve the effectiveness of code review comment generation task.
For example,
Hong et al. \cite{hong2025retrieval} proposed to retrieve review comments corresponding to code snippets that are most similar to the given input code.
Similarly, Olewick et al. \cite{olewicki2024impact} proposed to retrieve related comment examples for a given code change.
These RAG-based approaches are typically designed for mature, long-standing software projects with substantial historical data. 
However, they are less suitable for newly created projects that lack sufficient historical context~\cite{pornprasit2024fine}.

\textbf{Finally, an enterprise-grade code review comment generation should generate valid and factually correct code review comments}. 
Defined by Tufano et al. \cite{tufano2024code}, valid review comments refer to comments that provide clear suggestions aimed at improving the source code quality.
Liu et al. \cite{liu2025too} found that many code review comments could be noisy comments (e.g., vague or non-actionable feedback).
For example, \emph{``Add a blank line here"}.
Such noisy comments may lead LLMs to generate low-quality review comments.
To make matters worse, LLMs are prone to hallucination~\cite{ji2023survey,zhang2025llm,huang2025survey}---i.e., generating code review comments that are inaccurate, inconsistent, or nonsensical based on the provided code change.
Zhang et al. \cite{zhang2025llm} found that such hallucination is commonly found in the code generation task.
Such invalid code review comments may diminish the overall utility of the enterprise-grade automated code review tools.
These practical challenges of designing an enterprise-graded code review comment generation lead us to formulate the ultimate research question: \textbf{\textit{How can we design a review-guided, context-aware, quality-checked code review comment generation without fine-tuning?}}

\section{Atlassian's RovoDev Code Reviewer}\label{sec:rovodev}
\label{sec:rovodev}

In this section, we present the technical architecture of Atlassian's RovoDev Code Reviewer.

\begin{figure*}[t]
\includegraphics[width=\linewidth]{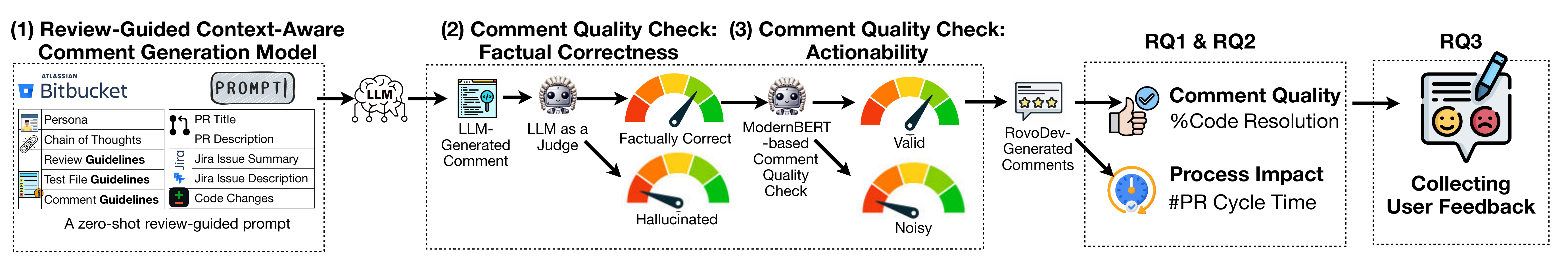}
\caption{An overview of our RovoDev Code Reviewer.}
\label{fig:overview}
\end{figure*}

\textbf{Overview.} We design RovoDev Code Reviewer as a zero-shot, review-guided, quality-checked, large language model (LLM)-driven approach to generate code review comments.
Given a pull request and its associated contextual information—such as the pull request title, description, relevant files, and linked issue summaries—the goal of RovoDev Code Reviewer is to automatically generate review comments that are valid, accurate, relevant, and actionable, and are precisely positioned within the appropriate locations in the code. 
Figure~\ref{fig:overview} presents an overview of our approach, consisting of the following three key components.

\textbf{\circled{1} A Zero-Shot Context-Aware Review-Guided  Comment Generation Model.} To ensure the privacy and security of our customers' data and maintain regulatory compliance, we do not collect any user-generated content for model training or fine-tuning purposes. 
Instead, we rely on a zero-shot structured prompting approach that is augmented with readily-available contextual information (i.e., pull request and Jira issue information) in a structured design with persona, chain-of-thought, and review guidelines.
With this approach, RovoDev Code Reviewer is more scalable to generate code review comments regardless of the maturity of the projects.

\textbf{\circled{2} A Comment Quality Check Component on Factual Correctness.} To ensure that code review comments being generated are not hallucinated (i.e., aligns with a given code change and its associated pull request), we use a LLM-as-a-Judge~\cite{zheng2023judging} approach to verify if a given comment is aligned with a given pull request and the area of code change or not.

\textbf{\circled{3} A Comment Quality Check Component on Actionability.} To ensure that code review comments being generated are valid and actionable (i.e., clear suggestions with a given code change), we propose a ModernBERT-based comment quality check to filter out noisy, vague, and non-actionable comments.
We describe each key component below.

\subsection{A Context-Aware Review-Guided Comment Generation}\label{sec:prompt}

Contextual information plays an important role in many tasks in software engineering~\cite{xu2020towards,wen2018context}, especially, code review automation~\cite{pornprasit2025context}.
Prior studies leverage various types of contextual information e.g., similar code changes to augment the code review comment generation task~\cite{hong2025retrieval, olewicki2024impact, hong2022commentfinder}.
However, such similar code changes require information access to within- or across-repositories, which may pose risks of data privacy and security and may not be available for the newly-created or context-limited software projects.

To address this challenge, RovoDev Code Reviewer operates in a zero-shot setting with Anthropic's \texttt{Claude 3.5 Sonnet}, leveraging only the provided context without relying on publicly-available dataset for training or fine-tuning, thereby upholding strict standards of data privacy and security. 
To ensure that the RovoDev-generated code review comments are valid, accurate, relevant, and actionable, and are precisely positioned within the appropriate locations in the code, we design our context-aware prompt that consists of the following key instructions and context information.

$\bullet$ \textbf{Persona} ($P$) is an explicit instruction to set the LLM to act as a software engineer, similar to prior studies~\cite{olewicki2024impact,pornprasit2024fine}.

$\bullet$ \textbf{Task Definition} ($T$) is an explicit instruction for LLM to perform a code review task for a given pull request.

$\bullet$ \textbf{Chain of Thoughts} ($C_T$) is a step-by-step reasoning process before generating a set of review comments. Our chain of thoughts provide concrete steps of code review tasks and what output format should be generated.

$\bullet$ \textbf{Review Guidelines for Code} ($G_\mathrm{Code}$) are explicit directives, well-crafted by Atlassian Engineering, provided to the LLM, ensuring that code reviews are conducted in strict alignment with the highest standards of enterprise quality and best practices (i.e., what should/should not be reviewed).

$\bullet$ \textbf{Review Guidelines for Test File } ($G_\mathrm{Test}$) are provided to ensure comprehensive evaluation of coverage, quality, and effectiveness of test files. 

$\bullet$ \textbf{Review Guidelines for Comment} ($G_\mathrm{Comment}$) are provided to guarantee that feedback is delivered with appropriate tone and structure, fostering constructive and professional communication throughout the review process.

$\bullet$ \textbf{Pull Request Title and Description} ($I_{PR}$) are provided as an additional context information to augment the prompt since it often gives high-level intent, motivation, and summary of the changes, which helps RovoDev Code Reviewer understand the purpose and scope of the given pull request.

$\bullet$ \textbf{Jira Issue Summary and Descriptions} ($I_{JIRA}$) typically contain the business motivation, requirements, and acceptance criteria for the given pull request.

$\bullet$ \textbf{Code Change} ($C_{code}$) is a set of code change for a given pull request.

Formally speaking, let ($\mathcal{P}=\{P, T, C_T\}$) denotes a system prompt, ($\mathcal{G}=\{G_R, G_{Test}, G_C\}$) the set of guidelines, ($\mathcal{I}=\{I_{PR}, I_{JIRA}\}$) the set of contextual information in a given pull request and Jira issue, and ($C_\mathrm{code}$) a code change; the code review generation task seeks a function ($f:(\mathcal{P}, \mathcal{G}, \mathcal{I}, C_{code}) \rightarrow R $), where ($R$) is a set of generated review comments with its associated file path and line number that maximize contextual relevance, factual correctness, and actionability, that eventually lead to a resolution of the pull request.


\subsection{Comment Quality Check on Factual Correctness}

It is possible that the set of generated code review comments ($R$) may be inaccurate, inconsistent, or nonsensical based on the provided code change, due to the LLM hallucination~\cite{ji2023survey,zhang2025llm,huang2025survey}.
To avoid LLM being hallucinated, we use LLM-as-a-Judge~\cite{li2024generation,zheng2023judging} ($J_\mathrm{fact}$) to evaluate if the set of generated code review comments ($R$) are factually correct (i.e., the generated comment is aligned with a given code change).
Our LLM Judge ($J_\mathrm{fact}$) is based on a prompting technique on a \texttt{gpt-4o-mini} model where our prompt consists of persona, task definition, instructions, and assessment guidelines.
Formally speaking, given an LLM Judge $J_\mathrm{fact}$, the assessment process can be formulated as $R=\{C_i,...,C_j\}$, where $C_i$ is the $i^{th}$ candidate to be judged and $R$ is the judging result.
We utilize a selection-based judgment methodology~\cite{li2024generation} that returns a binary outcome (True/False) in order to select one or more valid comment candidates to avoid the subjective threshold selection that is commonly presented in the score-based and ranking-based judgment.
Therefore, only code review comments that are factually correct will be proceed to the next component.

\subsection{Comment Quality Check on Actionability}

Although the generated-code review comments are factually correct, they may be vague (e.g., \emph{``Needs improvement"}), nitpicking without context (e.g., \emph{``Add a blank line here"}), unfocused or off-topic comments (e.g., \emph{``I saw something similar in another project"}), comments lacking specificity (e.g., \emph{``Is this the best way?"}), and non-actionable comments (e.g., \emph{``Good job!"}).
To address this challenge, we develop a ModernBERT-based Comment Quality Gate model to filter out such noisy code review comments generated by LLMs.
ModernBERT~\cite{warner2024smarter} is a modernized version of the encoder-only BERT model trained on 2 trillion tokens with a native 8,192 sequence length, allowing a longer context length for our Comment Quality Check component to infer if a given comment is of high quality or not.
To this end, we develop our Comment Quality Check model to capture the characteristics of RovoDev-generated code review comments that are most likely to drive code resolution. 
We fine-tuned our ModernBERT model using five months of over 50,000 high-quality RovoDev-generated code review comments from Atlassian’s proprietary repositories.
Note that we only use the historical data for model training to avoid data leakage in the online experiment.
Each data point consists of a RovoDev-generated comment paired with a label indicating whether it led to a code resolution ($<$comment, resolved?$>$).
Therefore, only code review comments that are valid and most likely to drive code resolution will be recommended to the given PR.

\subsection{A Seamless Integration of RovoDev Code Reviewer into Atlassian Bitbucket}

Following our prior study~\cite{takerngsaksiri2024human} and Atlassian's Human-AI Collaboration Strategy as the Future of Teamwork, a human-in-the-loop paradigm is maintained. 
Rather than automating the whole code review process, RovoDev Code Reviewer serves as a code review assistant, aiming to generate review comments and suggestions, but final decisions remain with human reviewers, supporting trust, accountability, and iterative improvement, and respecting human autonomy.
To ensure the scalability of our RovoDev Code Reviewer, we use an Event-driven Architecture (EDA), i.e., a software design paradigm that determines the flow of RovoDev Code Reviewer system by events changes.
Briefly speaking, when a pull request is created, RovoDev is triggered to begin the automated code review process as follows:

Step \circled{1}: RovoDev Code Reviewer clones a given repository and gathers all relevant information, including the code diff, PR title, PR description, and Jira issues \& descriptions.

Step \circled{2}: The collected context information is processed and fed into a large language model (LLM), which generates a set of candidate code review comments. These comments can include suggestions, bug detections, style improvements, and even suggested code changes that users can directly apply.

Step \circled{3}: RovoDev Code Reviewer applies a comment quality check following the overview workflow described in Section~\ref{sec:rovodev} (see Figure~\ref{fig:overview}) to  to select factually correct and actionable code review comments.

Step \circled{4}: The selected and validated comments are posted directly to the pull request, where human reviewers can review, accept, or further discuss them. 
Figure~\ref{fig:rovodevcodereviewer} presents an example of a code review comment generated by RovoDev Code Reviewer.

\section{Research Methodology and Results}\label{sec:experiment}

In this section, we present the online production steps, the motivation, research methodology, and results to answer our three research questions.


\begin{figure*}[t]
    \centering
    \begin{subfigure}[t]{\linewidth}
        \centering
        \includegraphics[width=.9\linewidth]{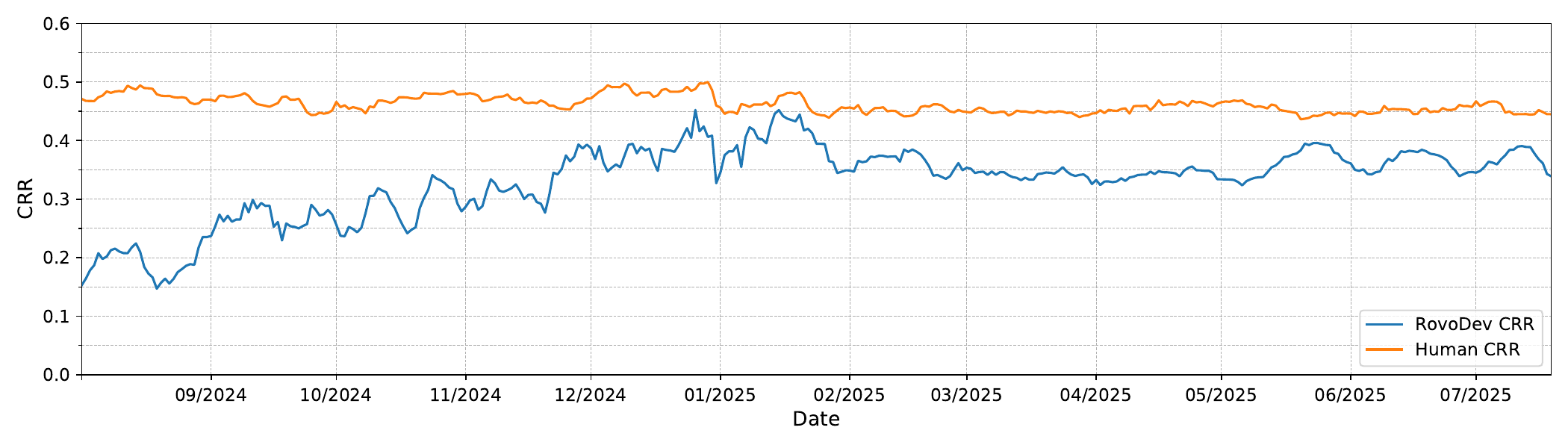}
        \caption{(RQ1) The code resolution rate of RovoDev-generated comments and human-written comments over a one-year period.}
        \label{fig:rq4}
    \end{subfigure}%
    \newline
    \begin{subfigure}[b]{0.49\linewidth}
        \centering
        \includegraphics[width=\columnwidth]{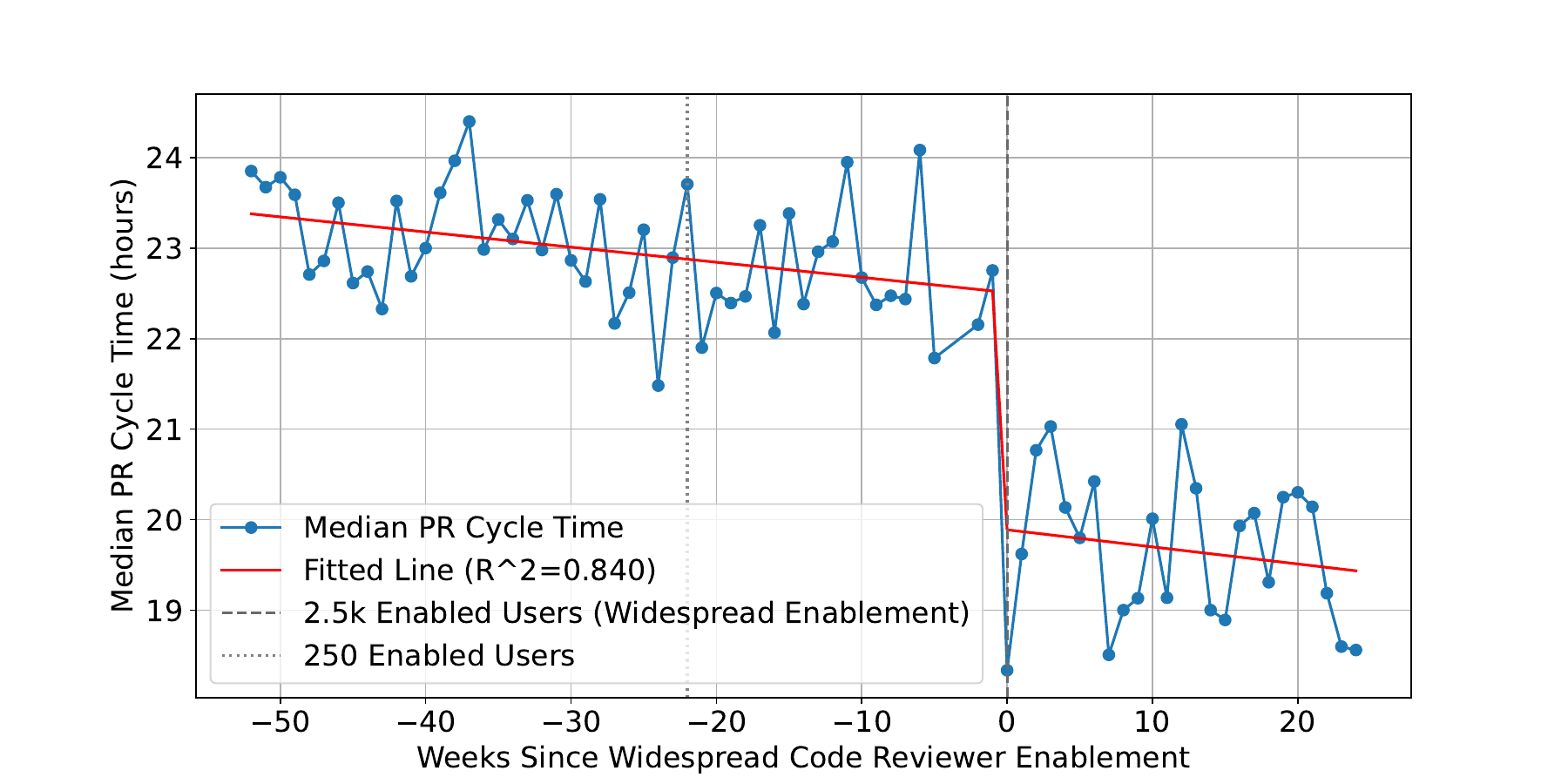}
        \caption{(RQ2) The median PR cycle time (hours) in Atlassian repositories over RovoDev Code Reviewer's adoption.}
        \label{fig:rq5-prepost-cycle}
    \end{subfigure}
    ~ 
    \begin{subfigure}[b]{0.49\linewidth}
        \centering
        \includegraphics[width=\columnwidth]{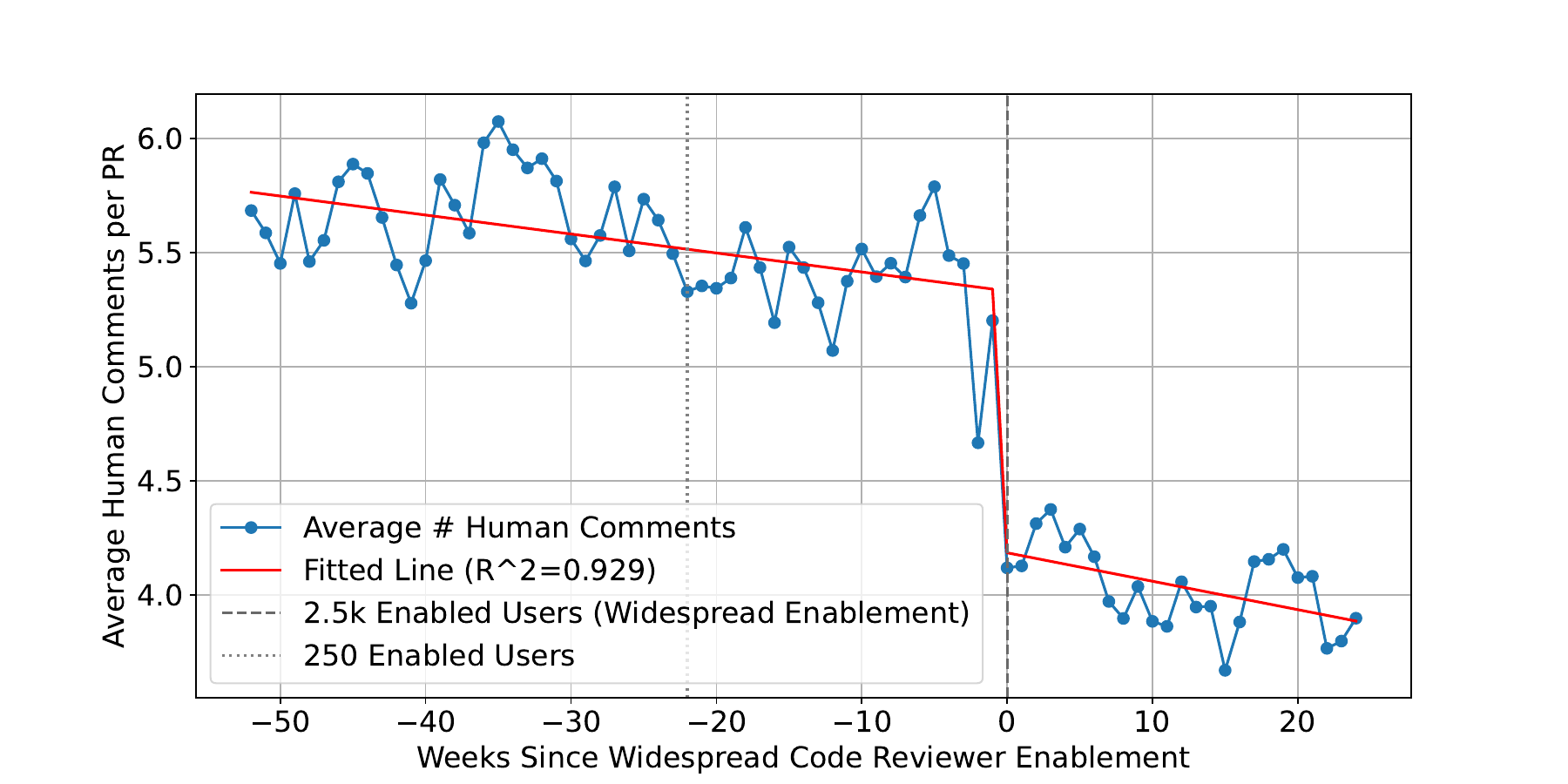}
        \caption{(RQ2) The average amount of human comments per PR in Atlassian repositories over RovoDev Code Reviewer's adoption.}
        \label{fig:rq5-prepost-comments}
    \end{subfigure}
    \caption{The results of the online evaluation stage for RovoDev Code Reviewer.}
\end{figure*}


\subsection{Online Deployment}

We deployed RovoDev Code Reviewer within Atlassian using the following roadmap.

\textbf{\circled{1}} \textbf{Internal Dogfooding (1 month).} To avoid interruption in the day-to-day work of Atlassian software engineers, we opted to deploy RovoDev Code Reviewer carefully as per the following steps. First, we conducted initial trials with a small group of software engineers from two software development teams to gain initial feedback. After receiving positive feedback from the first group, we gradually deployed RovoDev Code Reviewer with a broader group of Atlassian software engineers. 
We collected feedback to gain deeper insights into user experiences and the quality of the generated code review comments. 
At that time, the major concern is mostly related to the  quality of code reviewer comments generated by LLMs.
We conducted several iterations of trials to stabilize the quality of our RovoDev Code Reviewer-generated comments.

\textbf{\circled{2}} \textbf{Internal General Availability (12 months).} After improving RovoDev Code Reviewer based on the internal dogfooding, we expanded the deployment to all Atlassian teams. 
Over a 12-month deployment period between June 2024 and June 2025, RovoDev Code Reviewer was deployed across over 2,000 source code repositories, generating over 54,000 code review comments, with an average of 2.1 RovoDev-generated comments per pull request.
Such RovoDev-generated comments have been accepted by over 5,500 software engineers.
    
    



\subsection{A Large-Scale Online Evaluation}

\underline{\textbf{Motivation}}. 
As is common in code review automation research, any offline evaluation often relies on human-written comments as ground truth~\cite{hong2025retrieval, frommgen2024resolving, olewicki2024impact, liu2025too, tufan2021towards, tufano2022using, tufano2024code, llamareviewer, pornprasit2024fine} with several inherent limitations.
First, many code review benchmark dataset is static~\cite{tufan2021towards, tufano2022using, tufano2024code, llamareviewer}, which does not capture the full complexity, variability, or evolving nature of real-world software engineering tasks.
Second, the commonly-used semantic matching evaluation metrics (e.g., an exact match, BLEU, semantic similarity)~\cite{liu2025too} is constrained by the similarity to the human-written comments, but does not adequately reflect the practical usefulness (i.e., whether the comment actually facilitates code resolution).
In reality, code review comments can be written differently, and can be uniquely different from the human-written comments, yet share the same or similar intents.
Therefore, an online evaluation is critically needed to rigorously investigate the practical value of code review comments generated by RovoDev Code Reviewer.

To answer RQ1 and RQ2, we evaluate the practical value of RovoDev Code Reviewer based on the following aspects. 

$\bullet$ \textbf{Practical Value.} We hypothesize that high quality comments (e.g., clear, relevant, and actionable) should lead to code modification by software engineers.
Therefore, we measure, \emph{\% Code Resolution Rate (CRR)}, defined as the percentage of RovoDev-generated comments where the commented code lines are subsequently changed and the issue is resolved by software engineers in the next commit (aka. code resolution). 
High code resolution rates indicate the higher quality of the RovoDev-generated comments (e.g., clear, relevant, and actionable), which is well accepted by software engineers.
    
    
$\bullet$ \textbf{Process Improvement.} A key promise of code review automation is to accelerate the review process and reduce bottlenecks, thereby improving engineering productivity and software delivery speed. 
Therefore, we measure, \emph{PR Cycle Time}, defined as the duration from when a pull request (PR) is created to when it is successfully merged into a production environment or reaches a build endpoint. Shorter PR cycle times indicate faster resolution of pull requests, and are often associated with higher team productivity and better development practices.

%



\subsection*{\textbf{RQ1: \rqfour}}

\smallsection{\underline{Results}} \textbf{RovoDev Code Reviewer achieves a code resolution rate of 38.7\%, which is 12.9\% less than the code resolution rate of human-written comments (44.45\%).}
Figure \ref{fig:rq4} presents the Code Resolution Rate (CRR) of RovoDev-generated comments and human-written comments over a one-year period (Aug 2024 to July 2025), using a rolling average of 7-day (i.e., an average CRR value within a moving window of 7-days).
This one-year period includes the beginning of RovoDev's internal dogfooding (Aug 2025) to the widespread enablement, until July 15 2025. 
When considering the recent one-week period between July 10 2025 and July 15 2025, RovoDev Code Reviewer achieves a code resolution rate of 38.70\%, while human-written comments achieves a code resolution rate of 44.45\%. 
This leads to a maximum difference of 5.8 percentage points, or a relative percent difference of just 12.9\% ($\frac{38.70-44.45}{44.45}$) from the code resolution rate of human-written comments, highlighting the high practical value and quality of RovoDev-generated code review comments.
In addition, the most resolved types of code review comments are related to readability, bugs, and maintainability \cite{goldman2025types}.

\subsection*{\textbf{RQ2: \rqfive}}
\smallsection{\underline{Results}} 
\textbf{RovoDev Code Reviewer has a significant positive impact on the code review workflow at Atlassian, accelerating the median PR cycle time by 31\%, and reducing the number of human-written comments by 36\%.}

\subsubsection*{RQ2.1) PR Cycle Time}

To measure the impact of RovoDev Code Reviewer on the code review workflow, we first look at the PR cycle time. We first make a comparison within enabled repositories of standard size ($\leq 50$ PRs merged per week, $\leq 25$ unique PR authors per week), over the period of widespread internal enablement (January 2025 through July 2025). We take the cycle time of PRs (hours) that have no RovoDev Code Reviewer comments, and compare the cycle time of PRs that have Code Reviewer comments. 
Table \ref{tab:pr_cycle_time} shows that, at the median value, RovoDev Code Reviewer can speed up the cycle time of PRs by 31\% ($\frac{14.35-20.73}{20.73}$) with a statistically significant difference (Mann-Whitney U test, $p<.001$).

\begin{table}[h]
\centering

\caption{(RQ2.1) The cycle time of pull requests (hours) between with and without RovoDev comments.}
\resizebox{\columnwidth}{!}{%
\begin{tabular}{|l|r|r|r|r|r|}
\hline
PR Type & \#PRs & Q1 & Median & Q3 \\
\hline
With RovoDev Comments & 43,633 & 1.06 & 14.35 & 47.67 \\
Without RovoDev Comments & 42,981 & 2.40 & 20.73 & 73.25 \\
\hline
\textbf{Relative Difference (\%)} & & \textbf{-56\%} & \textbf{-31\%} & \textbf{-35\%} \\
\hline
\end{tabular}}
\label{tab:pr_cycle_time}
\end{table}

In order to confirm this result is not due to sample bias, we then calculate RovoDev Code Reviewer's impact on Atlassian's PR cycle time immediately after widespread enablement. Figure~\ref{fig:rq5-prepost-cycle} shows the median PR cycle time in enabled Atlassian repositories per week. It covers 77 weekly observations, spanning from 52 weeks before to 24 weeks after Code Reviewer's widespread internal enablement. From these observations, we perform an interrupted time series analysis using ordinary least squares (OLS) regression to evaluate the impact of RovoDev Code Reviewer on the PR cycle time, as shown in the red line in Figure~\ref{fig:rq5-prepost-cycle}. The model includes a binary intervention variable indicating the introduction of the RovoDev Code Reviewer (i.e., at week 0).
The analysis reveals a statistically significant reduction in the median pull request cycle time following the widespread internal enablement of RovoDev Code Reviewer.

\subsubsection*{RQ2.2) \#Human Comments per PR}
We also measure the impact of RovoDev Code Reviewer on the code review workflow by looking at the average number of human comments per PR, a measure of the extent of human effort and intervention in the code review process. We take the same set of PRs, coming from standard-sized enabled repos over the same period (including and after widespread enablement). Table \ref{tab:human_comments_count} shows that, at the median value, RovoDev Code Reviewer can reduce the number of human-written comments by 35.6\% ($\frac{2.87-4.45}{4.45}$) with a statistically significant difference (Mann-Whitney U test, $p<.001$).



\begin{table}[h]
\centering
\caption{(RQ2.2) The number of human-written comments per PR between with and without RovoDev comments.}
\begin{tabular}{|l|r|r|}
\hline
PR Type & \#PRs & Mean \\
\hline
With RovoDev Comments & 43,633 & 2.87 \\
Without RovoDev Comments & 42,981 & 4.45 \\
\hline
\textbf{Relative Difference (\%)} & & \textbf{-35.6} \\
\hline
\end{tabular}

\label{tab:human_comments_count}
\end{table}


Figure~\ref{fig:rq5-prepost-comments} shows the average of human comments per PR over Atlassian's repositories per week. As with the PR cycle time analysis, it covers 77 weeks of observations, 52 weeks before and 24 weeks after widespread enablement. We perform the same interrupted time series analysis using OLS, using the same binary intervention variable.

With a similar trend to the PR cycle time, this analysis showed a statistically significant reduction in the mean amount human comments per PR after RovoDev Code Reviewer was widely enabled $(p = 0.013)$. This is an indicator that the widespread introduction of RovoDev Code Reviewer correlated with a significant boost in engineers' productivity.

With both the PR cohort analysis and interrupted time series analysis showing statistically significant reductions in both PR cycle time and human comments per PR, these findings confirm that RovoDev Code Reviewer can speed up review cycle, while alleviating reviewer workload.

\subsection{\textbf{User Feedback Evaluation.}}


\textbf{\underline{Motivation.}} User evaluation feedback is a critical component in both academic research and product development, as it provides direct insights into user needs, behaviors, and experiences. 
However, there exist qualitative research methods such as survey and interview used for evaluating automated code review in practice. 
For example, Cihan et al.~\cite{cihan2024automated} conducted a board survey with 22 practitioners to investigate the practitioner's perception on automated code review in practice.
However, such survey studies only capture the general opinions from software engineers rather than the specific fine-grained qualitative feedback at each individual comment (i.e., why a particular code review comment is useful or not useful).


\textbf{\underline{Research Methodology.}} To address the gap, we collected feedback at each individual comment.
For each comment generated by RovoDev, we explicitly ask software engineers to rate the comment quality using the thumbs-up or thumbs-down.
However, similar to a case study at  Google~\cite{vijayvergiya2024ai}, we observe that software engineers rarely clicked the thumbs up/thumbs down buttons.
Therefore, we also ask the software engineers to provide an explicit feedback if they find a RovoDev-generated comment particularly useful or not with an option for our RovoDev team to make a direct contact with the software engineers or not, respecting user privacy.
All of the comments were collected and fed into the Atlassian's Jira Service Management Cloud System for feedback analysis, prioritization, and management.
We conducted a thematic analysis to identify the common themes of the feedback data using an open-coding approach.
Following a Reflexive Thematic Analysis (RTA)~\cite{byrne2022worked}, we manually reviewed each feedback and assigned a code and theme that is related to the feedback.
With the reflexive approach, codes and themes will be evolved throughout the process, meaning that there is no fixed codebook and boundaries of codes can be redrawn, split, merged, or promoted to themes as understanding deepens.
For example, \emph{``This is an amazing catch''} and \emph{``Nope we were wrong, the AI was right!''} were grouped into one theme.

\subsection*{\textbf{RQ3: \rqsix}}

\smallsection{\underline{Results}} Through a collection of qualitative feedback collected over the 12-month deployment period, we present the two major feedback themes as follows.

\textbf{Accurate Error Detection and Actionable Suggestions.} Software engineers found that RovoDev Code Reviewer is highly valued for its ability to accurately detect subtle errors and redundancies in code (e.g., a duplicate method name), providing actionable suggestions that enhance code quality and support efficient development workflows.
For example, users mentioned that \emph{``This was helpful. Basically, it caught a typo which could have broken the functionality''}, \emph{``The CodeReviewer made a helpful comment how to improve my Jira expression in a Forge manifest!}.
On the other hand, a user found that RovoDev Code Reviewer can provide good suggestion yet incorrect, \emph{``Good suggestion. Incorrect. But highlighted something worth commenting in the code itself.''}.

\textbf{Lack of holistic code understanding.} Software engineers suggest that comments generated by RovoDev Code Reviewer could be more specific and accurate by improving its holistic understanding of the source code context. 
For example, \emph{``This is Jira Expression where $\mathrm{!==}$ is not acceptable''}, \emph{``AI seems to think this is PHP code, it's Javascript''}.
These feedback highlights that RovoDev Code Reviewer still needs additional syntactic information and programming environments (e.g., programming languages, frameworks, versions, libraries, naming and coding standards, deprecated API calls, code structure, program dependency, and surrounding code context). 
However, incorporating all of the possible context information into RovoDev is nearly infeasible due to the limited context window of LLMs (i.e., the amount of text or tokens that a model can consider at once when generating a comment for a given pull request). 
Therefore, researchers should consider investigating the best context enrichment approach for code review comment generation.

\section{Discussion \& Recommendation}\label{sec:discussion}

In this section, we conduct additional analysis to investigate the alignment between RovoDev-generated comments and human-written comments based on our internal benchmark dataset, the impact of our prompt component, and the impact of the comment quality check on the effectiveness of RovoDev Code Reviewer.
Finally, we draw recommendations based on results and lessons learned.

\subsection{\large{\rqone}}

\textbf{\underline{Motivation.}}
The alignment of LLM-generated comments to human-written comments is crucial for several reasons. 
Human-written comments encapsulate domain expertise, organizational standards, and nuanced understanding of code context, which are essential for effective feedback and knowledge transfer. 
Therefore, achieving a high degree of alignment between LLM-generated and human-written code review comments is not only desirable but necessary for the practical deployment of automated code review tools in real-world code review practices.

\textbf{\underline{Dataset.}}
We curated our own code review benchmark dataset from Atlassian's proprietary software repositories.
However, not all code changes have associated PR and Jira information.
Our initial dataset consists of 3,492 code changes and their associated high-quality, human-written code review comments.
To ensure that the experiment is in a controlled setting, we filtered out the code changes that do not have pull request and Jira information.
Thus, we removed 872 code changes that do not have both associated PR and Jira information.
Prior studies shown that many comments written by human could be noisy (e.g., \emph{``Looks good to me''}, or \emph{``/nit the space!"''})~\cite{liu2025too}.
Following Liu et al~\cite{liu2025too}, we exploited an LLM to filter out 270 code changes that are associated with noisy human-written comments (i.e., the affirmation and humour types of comments).
Code review practices have matured over time, and recent comments reflect the latest standards and methodologies. 
To ensure that the comments are aligned with current best practices and organizational goals, we focused on the code review comments written by human reviewers over the past 3 years (2022-2025), removing 282 code changes.
In summary, our internal benchmark dataset consists of 2,068 code changes that are associated 2,894 high-quality, human-written code review comments, spanning across 1,468 PRs.

\underline{\textbf{Evaluation Measures.}}
We consider the following two evaluation aspects.
\emph{Human-aligned Review Localization} refers to the capability to generate comments at locations that closely correspond to where a human reviewer would naturally place them~\cite{hong2022commentfinder, hong2022should}, while \emph{Human-aligned Comment Generation} focuses on the capability to generate comments that is similar to human-written comments~\cite{hong2025retrieval, xu2020towards}.
In the ideal scenario, RovoDev-generated comments should be at the same location and textually-similar as human-written comments.
Therefore, we measure:


\textbf{\%HAC (Human-Aligned Comments, capturing both location \& semantic similarity)} measures the percentage of human-aligned comments out of all comments generated by RovoDev. 
We consider a code review comment to be \emph{human-aligned} if it meets two conditions: (1) it is at a close-proximity location (i.e., at exactly the same file path and within a $\pm$10 line range of the actual location); and (2) it is semantically highly similar to the human-written comment (i.e., similar comment intent).
To compute the semantic similarity, we use LLM-as-a-Judge~\cite{li2024generation,zheng2023judging} ($J_\mathrm{sim}$) to evaluate if the set of generated code review comments ($R$) are semantically similar to a given human-written comment.
Similar to Section 3.2, our LLM Judge ($J_\mathrm{sim}$) is based on a prompting technique using a \texttt{gpt-4o-mini} model where our prompt consists of persona, task definition, instructions, and assessment guidelines.
We employ a score-based evaluation method~\cite{li2024generation} that assigns a similarity score to each pair of code review comments: 1 (no similarity), 2 (weak similarity), 3 (high similarity), and 4 (very high similarity). 
Pairs receiving a score of 3 or 4 are considered to exhibit high similarity to the human-written comment.
Nevertheless, it is possible that RovoDev may generate comments that are not aligned with human.
Thus, we measure \textbf{\%!HAC (Human-Not-Aligned Comments)} as the percentage of RovoDev-generated comments that are not aligned with the human-written comments.
Since RovoDev Code Reviewer is operated at the pull request level, to gain better insights, we also operationalize the following metrics at the pull request level.

\begin{itemize}
    \item \textbf{\%PR\_{HAC}} measures the percentage of pull requests that contains at least one human-aligned RovoDev-generated comment (i.e., both at the correct location and semantically similar to the human-written ground-truth comment). 
    \item \textbf{\%PR\_{!HAC}} measures the percentage of pull requests that contains at least one RovoDev-generated comment that is not aligned with the human-written comment for a given pull request. 
\end{itemize}

To ensure that the results are robust and reliable, we repeat the experiment 5 times and present the bar chart of the mean value and its 95\% confidence interval.

\begin{figure}[t]
\centering
\includegraphics[width=\columnwidth]{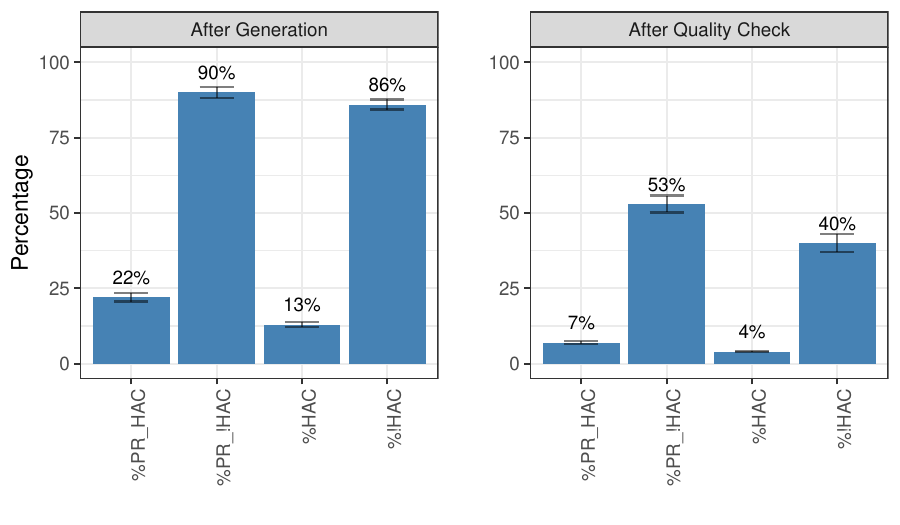}
\caption{An evaluation of LLM-human comment alignment, measured by \%HAC (\% of RovoDev-generated comments that are aligned with human-written comments, capturing both location and semantic similarity).}
\label{fig:rq1}
\end{figure}


\textbf{\underline{Results}} 
\textbf{With the end-to-end evaluation workflow of RovoDev Code Reviewer, there are 7\% (\%PR\_HAC) of the pull requests that contain at least one comment that is aligned with a human-written comment.}
Figure~\ref{fig:rq1}-\emph{right} shows that, after the quality check stage, RovoDev Code Reviewer can generate 4\% of the comments that are aligned with human-written comments, but still producing a moderately high number (40\%) of comments that are not aligned with human-written comments.
Nevertheless, it is important to note that a lack of alignment between RovoDev-generated and human-written comments does not necessarily imply that RovoDev's comments are not useful as RovoDev and humans may focus differently.
Therefore, the RovoDev-generated comments are still valuable, since our RQ1 shows that RovoDev Code Reviewer achieves a high code resolution rate (38.70\%) (i.e., generate actionable comments that trigger a code resolution). 

From the industrial perspectives, code review comments can be written in many different forms, regardless of their intended purpose. 
Traditional methods for evaluating these comments often depend on similarity scores (such as BLEU) or exact matches~\cite{tufano2022using, tufano2021towards,liu2025too}, which restrict their ability to truly measure the quality and relevance of the generated code review comment. 
As a result, valuable insights that are expressed differently but share the same intent may be missed, leading to an incomplete evaluation of a given code review comment. 

\underline{\textbf{Recommendation.}} We recommend that the similarity metric should not be used as the sole measure of the practical usefulness of such comments. Instead, to fully assess the impact and value of automated code review, an online evaluation is necessary.

\begin{figure}[t]
\centering
\includegraphics[width=\columnwidth]{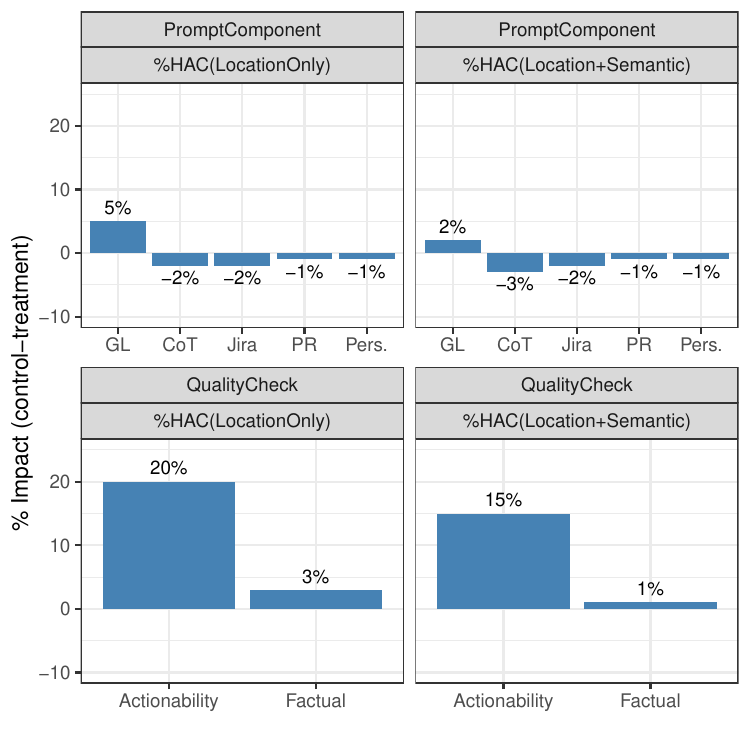}
\caption{The impact of the prompt components and comment quality check on the effectiveness of RovoDev Code Reviewer. The impact is measured by the absolute percentage difference (\%$_\mathrm{control}$ - \%$_\mathrm{treatment}$).}
\label{fig:rq23}
\end{figure}

\subsection{\large{Which types of prompt components are most important for RovoDev Code Reviewer?}}


\textbf{\underline{Motivation.}} 
Our zero-shot context-aware prompt is structured, yet consists of multiple components (see Section~\ref{sec:prompt}).
To gain a better understanding of which types of prompt components are most valuable, we set out to investgate the impact of the prompt components on the effectiveness of RovoDev Code Reviewer.

\textbf{\underline{Approach.}} 
To do so, we compare the effectiveness of RovoDev Code Reviewer between the full prompt version (\emph{a control group}) and the five prompt variations (\emph{a treatment group}, i.e., a  prompt where each individual component is removed).
We consider that Task Definition ($T$) and Code Change ($C_{code}$) are mandatory information, which must be present in the system prompt.
For the five prompt variations, we consider removing the persona ($P$), the chain of thoughts ($C_T$), the set of guidelines ($\mathcal{G})$, the PR information ($I_{PR}$), and the Jira issue information ($I_{Jira}$) from the full prompt version.

\textbf{\underline{Results.}} 
\textbf{Review Guideline together with Task Definition and Code Changes are the most important prompt components for code review comment generation, as it improves RovoDev's review localization capability by 5 percentage points.}
Figure~\ref{fig:rq23} presents the percentage impact of the prompt components on RovoDev's effectiveness. 
Positive impact indicates the positive benefit that a component contributes to RovoDev (i.e., an absolute percentage loss when such the treatment component is removed from the control group).
The results show that the review guideline contributes 5\% positively to the RovoDev's review localization.
In other words, when the review guideline is removed from the control group, \%HAC(LocationOnly) of RovoDev is decreased by 5 percentage points.
Nevertheless, other types of contextual information (e.g., persona, Chain-of-Thought, PR information, and Jira Information) have a minimal impact on the RovoDev's review localization and generation capability---the absolute percentage difference is only 1\%-3\%.
For Jira issues and pull requests, our deeper investigation found that such information is generally high-level and lack sufficient detail to effectively generate code review comments.
For persona, this finding is similar to the literature~\cite{pornprasit2024fine}---i.e., persona does not have much impact on the code review automation.

\underline{\textbf{Recommendation.}} We recommend that review guidelines, along with task definitions and code changes, be provided as mandatory inputs for zero-shot code review comment generation.

\subsection{\large{Which comment quality check components are most effective?}}

\textbf{\underline{Motivation.}} 
To maximize the practical utility of automated code review, it is essential that generated comments are both factually correct and actionable. 
To this end, we introduce a comment quality check component into the RovoDev Code Reviewer, focusing specifically on verifying factual correctness and actionability. 
Despite the intuitive benefits of such quality checks, there is limited empirical understanding of the relative impact of each component—factual correctness versus actionability—on the overall effectiveness of automated code review. 
Understanding which aspect contributes more significantly can inform the design of more effective automated review tools and guide future research in this area.

\textbf{\underline{Approach.}} 
To investigate the impact of comment quality check components on the effectiveness of RovoDev Code Reviewer, we conduct a controlled experimental study. 
We systematically compare the effectiveness of RovoDev Code Reviewer under different settings: (1) with both factual correctness and actionability checks enabled  (\emph{control}), (2) without actionability check enabled, and (3) without factual correctness check enabled (\emph{treatment}). 

\textbf{\underline{Results.}} 
\textbf{The Actionability Check helps RovoDev to generate 15\% more comments with correct locations, while the Factual Correctness Check has a minimal impact on the RovoDev's overall effectiveness.}
Figure~\ref{fig:rq23} shows that the ModernBERT-based Actionability Check helps improving the \%HAC(LocationOnly) by 20 percentage points, and improving \%HAC(Location+Semantic) by 15 percentage points.
On the other hand, the Factual Correctness based on the LLM-as-a-Judge, which tends to be expensive, has a minimal impact on the RovoDev's overall effectiveness.
This finding is surprising as it contradicts the findings in the ML literature, who found that such the zero-resource black-box Judge can reduce hallucination~\cite{manakul2023selfcheckgpt,chern2023factool}.

\underline{\textbf{Recommendation.}} We recommend prioritizing actionability checks based on pre-trained language models, such as the ModernBERT-based Actionability Check, as they substantially improve the human-alignment of RovoDev-generated comments. 

\section{Threats to the Validity}\label{sec:threats}

In this section, we disclose the threats to the validity.

\textbf{Threats to construct validity.} 
To evaluate the human alignment of our RovoDev-generated comments, we use LLM-as-a-Judge~\cite{zheng2023judging} as a proxy to measure the semantic similarity of a given code review comment to the associated human-written comment. 
Nevertheless, it is crucial that the Judge itself is highly reliable to ensure robust and trustworthy evaluation results~\cite{gu2024survey}.
To evaluate the reliability of our Judge, we collected a small sample of 47 human-written comments.
Then, we compared our Judge (\texttt{gpt-4o-mini}) with other models.
Results show that our Judge produces scores that are highly correlated with human judgement (a Spearman correlation of 0.69), outperforming other LLMs (0.65 for mixtral-8-7b, 0.57 for llama3.1-8b, 0.54 for llmma3.0) and other cosine similarity scores (<0.52), indicating that our Judge is of reliable enough.


\textbf{Threats to external validity.} Our RovoDev Code Reviewer approach is generally applicable to any LLMs (e.g., GPT4.1, Claude, Llama, Qwen, etc). 
However, the results of this paper are limited to Antropic Claude Sonnet 3.5 and the context of Atlassian's internal code review.
Therefore, our results may not be generalized to other LLMs and code review platforms. 



\section{Conclusion}\label{sec:conclusion}


In this paper, we present RovoDev Code Reviewer, a review-guided, quality-checked code review automation approach. 
Through a large-scale online evaluation involving over 1,900 Atlassian repositories and more than 54,000 generated code review comments over a year, we find that: (1) RovoDev Code Reviewer is effective in facilitating code resolution, with 38.70\% of its comments leading to code changes in subsequent commits, demonstrating its practical value in real-world settings; (2) RovoDev significantly improves the code review workflow at Atlassian, reducing median pull request cycle time by 30.8\% and decreasing the number of human-written comments by 35.6\%; and (3) while RovoDev can produce accurate and actionable comments, it may generate incorrect or non-actionable feedback when contextual information is lacking (e.g., unknown programming languages, versions, or frameworks). 
Notably, augmenting prompts with additional context yields only minimal improvements. These findings highlight the need for further research into advanced context augmentation techniques for LLM-powered code review automation.

\textbf{Disclaimer.}
The perspectives and conclusions presented in this paper are solely the authors' and should not be interpreted as representing the official policies or endorsements of Atlassian or any of its subsidiaries and affiliates. Additionally, the outcomes of this paper are independent of, and should not be constructed as an assessment of, the quality of products offered by Atlassian.


\bibliographystyle{ACM-Reference-Format}

{
\small
\bibliography{references}
}


\end{document}